\documentclass[10pt]{article}
\usepackage{amsfonts}
\usepackage{fullpage,graphicx,subfigure,mathdots,mathpazo,color}
\usepackage{amsmath,amscd,tikz,mathrsfs,cite}
\usepackage[normalem]{ulem}
\usepackage{amsmath}
\usepackage{setspace}
\usepackage{epsfig,amsmath,graphicx,amssymb,overpic}
\usepackage{bm}
\usepackage{graphicx}
\usepackage{subfigure, xcolor}
\usepackage{ntheorem}
\usepackage{listings,diagbox}
\usepackage{color}
\usepackage{epsfig,amsmath,graphicx,amssymb,overpic}
\usepackage{booktabs}
\usepackage{diagbox}
\usepackage{graphicx}
\usepackage{dcolumn}
\usepackage{bm}
\usepackage{graphicx}
\usepackage{subfigure}
\usepackage{epsfig,amsmath,graphicx,amssymb,overpic,cite}
\usepackage{makecell}
\usepackage{geometry}

\def\be{\begin{equation}}
\def\ee{\end{equation}}
\def\bee{\begin{eqnarray}}
\def\ene{\end{eqnarray}}
\def\bes{\begin{subequations}}
\def\ees{\end{subequations}}
\def\no{\nonumber}

\def\v{\vspace{0.1in}}
\def\no{{\nonumber}}

\allowdisplaybreaks[4]

\begin{document}

\baselineskip=14pt \renewcommand {\thefootnote}{\dag}
\renewcommand
{\thefootnote}{\ddag} \renewcommand {\thefootnote}{ }

\pagestyle{plain}

\begin{center}
\baselineskip=16pt \leftline{} \vspace{-.3in} {\Large \textbf{{
Suppression of soliton collapses, modulational instability, and rogue-wave excitation
%\\[0pt]
in two-L\'{e}vy-index fractional Kerr media}}} \\[0.1in]

Ming Zhong,$^{1,2}$\,\, Yong Chen,$^{3}$\,\, Zhenya Yan$^{{1,2},{*}}$\footnote{$^{*}$Corresponding
author. \textit{Email address}: zyyan@mmrc.iss.ac.cn} and Boris
A. Malomed$^{4,5}$ \\[0.1in]
\textit{{\small $^{1}$KLMM, Academy of Mathematics and Systems Science,
Chinese Academy of Sciences, Beijing 100190, China \\[0pt]
$^{2}$School of Mathematical Sciences, University of Chinese Academy of
Sciences, Beijing 100049, China \\[0pt]
$^{3}$School of Mathematics and Statistics, Jiangsu Normal University, Xuzhou 221116, China
\\[0pt]
$^{4}$Department of Physical Electronics, School of Electrical
Engineering, Faculty of Engineering,\\[0pt]
Tel Aviv University, Tel Aviv 69978, Israel \\[-4.5pt]
$^{5}$Instituto de Alta Investigaci\'{o}n, Universidad de Tarapac\'{a},
Casilla 7D, Arica, Chile}} % (Date:\,\, \today)
\end{center}

\vspace{0.1in}

{\baselineskip=13pt

%{\small \baselineskip=13pt \vspace{0.1in} %
\noindent {\bf Abstract:}\, We introduce a generalized fractional nonlinear Schr\"{o}dinger  equation for the
propagation of optical pulses in laser systems with two fractional-dispersion/diffraction terms, quantified by their L\'{e}vy
indices, $\alpha _{1}\, \alpha _{2}\in (1, 2]$,  and self-focusing or defocusing Kerr nonlinearity. Some fundamental
solitons are obtained by means of the variational approximation, which
are verified by comparison with numerical results. We find that the soliton collapse,
exhibited by the one-dimensional cubic fractional nonlinear Schr\"{o}dinger equation with only one L\'{e}vy index
$\alpha =1$, can
be suppressed in the two-L\'{e}vy-index fractional nonlinear Schr\"{o}dinger system. Stability of the solitons is
also explored against collisions with Gaussian pulses and adiabatic
variation of the system parameters. Modulation instability  of
continuous waves is investigated in the two-L\'{e}vy-index system too. In particular, the modulation instability may occur in the case of the defocusing nonlinearity when two diffraction coefficients
have opposite signs. Using results for the modulation instability, we produce first- and
second-order rogue waves on top of continuous waves, for both signs of the Kerr nonlinearity.
%\vspace{5pt}

%\vspace{-0.05in} %\begin{tabular}{p{16cm}}
%  \hline \\
%\end{tabular}The nonlinear properties of such systems are then determined by the  band gap and the type of nonlinearity.

%\vspace{0.1in}

\section{Introduction}

Fractional quantum mechanics, which was first proposed by Laskin~\cite{La00} in
2000, is based on the Feynman path integration performed over Brownian
trajectories replaced by L\'{e}vy flights. The derivation gives rise to the
linear fractional Schr\"{o}dinger equation (FSE)~\cite{La02}:
\begin{equation}
i\hbar \psi_t=D_{\alpha }\big(-\hbar
^{2}\nabla^2 \big)^{\alpha /2}\psi+V(t, \mathbf{r})\psi
,\,\,\,\,\,\, D_{\alpha }\in \mathbb{R},  \label{FLS}
\end{equation}%
where the fractional diffraction is characterized by the L\'{e}vy index (LI)
\cite{Mandelbrot}, which normally takes values in the interval
%\begin{equation}
$\alpha \in (1,2].$
%  \label{interval}\end{equation}%
The fractional kinetic-energy operator $(-\hbar ^{2}\nabla^2 )^{\alpha /2}$ with $\nabla^2$
being the Laplacian operator and $\hbar$ the reduced Planck constant in
Eq. (\ref{FLS}) is based on the Riesz fractional derivatives~\cite{riesz}
[the explicit definition is given below in Eq. (\ref{FT})], and $V(t,\mathbf{r})$
is the external potential. Equation (\ref{FLS}) amounts to the usual
linear Schr\"{o}dinger equation in the case of LI $\alpha =2$. While
experimental realization of the fractional quantum mechanics is not known,
the realization of the FSE, with time $t$
replaced by propagation distance $z$ and $x$ being the transverse coordinate
in the waveguide, was proposed in optics, using the similarity between the
quantum-mechanical Schr\"{o}dinger equation and propagation equation for the
optical amplitude under the action of the paraxial diffraction~\cite{Lo15}.
In the framework of this mechanism, the spatial fractional diffraction may
be implemented by means of the corresponding term in the Fourier space~\cite%
{La00,Lo15}. Another experimental realizations of the fractional diffraction
have been also proposed in condensed matter~\cite{St13,Pi15}. More recently,
Liu {\it et al}~\cite{Li23} first reported the implementation of the fractional group-velocity dispersion (GVD) in
the temporal domain (rather than diffraction in the spatial propagation) in experiments with fiber lasers,
where the corresponding
optical medium is modeled by the linear FSE with LI $\alpha \in (1,2]$ in the temporal domain,
\begin{equation}
i\psi_z=\Big[D_{\alpha}\left(-\partial_\tau^{2}\right) ^{\!\!\alpha /2}-\sum_{s=2,3,...}
\beta _{s}/s!\left( i\partial_\tau\right)^{s}\Big]\psi+W(\tau)\psi,\quad \partial_\tau=\partial/\partial \tau,
\label{FSE-esp}
\end{equation}%
where $D_{\alpha}$ is a real-valued fractional-dispersion parameter, $\beta _{s}$ the real-valued $s$-th regular GVD parameter, and $W(\tau)$ an effective potential. The setup decomposes the temporal optical pulse into its spectral
components, making them spatially separated. Each component, carried by its
wavelength, passes a dedicated segment of the phase plate and thus receives
a phase shift which emulates the expected contribution from fractional GVD
for the particular wavelength. Then, the separated components are recombined
back into the temporal pulse. The central element of the setup, \textit{viz}%
., the properly profiled phase plate, was created as a computer-generated
hologram. Essentially the same technique may be applied to the generation of
the effective fractional diffraction in the spatial domain.

In the scaled form, the evolution of the optical amplitude $\psi (\mathbf{r},z)$ in a
nonlinear waveguide with fractional diffraction is governed by the higher-dimensional
fractional nonlinear Schr\"{o}dinger (FNLS) equation (see, e.g., Ref.~\cite{FNLS9} and reference therein)
\begin{equation}
i\psi_z=\left( -\nabla^2 \right) ^{\alpha/2} \psi +U(\mathbf{r},z) \psi -g|\psi |^{2}\psi,  \label{FNLS-0}
\end{equation}%
where $g=+1$ and $-1$ represent, respectively, the cubic
self-phase-modulation term with self-focusing and defocusing signs,
and $U(\mathbf{r},z)$ is an effective potential, which may be induced by local modulation
of the refractive index in the waveguide. Various species of solitons were
predicted in the framework of FNLS equations, including spatiotemporal
\textquotedblleft accessible solitons"~\cite{FNLS10}, gap~\cite%
{FNLS12,FNLS13} and multi-pole modes~\cite{FNLS14,FNLS15}, and solitary
vortices in multi-dimensional settings~\cite{FNLS17,FNLS19}. A noteworthy
finding is that, when symmetry breaking occurs in FNLS equations, it may
give rise to asymmetric solitons with complex propagation constants~\cite%
{FNLS20,li-op21,FNLS21,FNLS23}.

Similar to the two-dimensional NLS equation~with the regular (non-fractional)
diffraction \cite{Ga66,Vl71}, the critical wave collapse (blow-up) of
configurations with the norm (total power, in terms of optics),%
\begin{equation}
P=\!\int_{-\infty }^{+\infty }\!\!|\phi (x)|^{2}dx,  \label{P}
\end{equation}%
exceeding a certain critical value,
\begin{equation}
P_{\mathrm{crit}}\approx 1.23,  \label{Pcrit}
\end{equation}%
has been predicted in the one-dimensional (1D) FNLS equation (\ref{FNLS-0})
with self-focusing ($g=+1$) and LI $\alpha =1$ \cite%
{Kl14,Ch18,Qiu20,Sakaguchi} (values $\alpha <1$ corresponds to the
supercritical collapse, which may be initiated by an arbitrarily small norm,
i.e., the respective critical power is zero). The wave collapse~\cite%
{Be98,Fi15} leads to the formation of a singularity after a finite
propagation distance (while an input with $P<P_{\mathrm{crit}}$ decays). The
wave collapse has been reported in plasmas~\cite{Wo84}, optics \cite{Gaeta},
Bose-Einstein condensates~(BECs) \cite{Sa99,Xie18}, capillary-gravity waves in
deep water~\cite{Ba83}, astrophysics~\cite{Ho96}, etc.

NLS equations which give rise to the critical collapse produce families of
\textit{Townes solitons} (TSs), which were first predicted in terms of the
2D\ cubic NLS equation \cite{Ga66}. These families are degenerate in the
sense that they exists with the single value of the norm [e.g., the one
given by Eq. (\ref{Pcrit}) for the 1D cubic\ FNLS equation with $\alpha =1$%
], and they are fully unstable \cite{Be98,Fi15}. The onset of the
instability is slow, as it initially develops subexponentially, which made
it recently possible to directly observe 2D TSs in binary BEC as
quasi-stable objects \cite{Bakkali}. Nevertheless, the critical collapse
eventually leads to destruction of TSs and emergence of the singularity. For
this reason, stabilization of TS-like solitons in physically relevant
settings is a problem of fundamental interest \cite{Mal22}.

A common way for the solution of this problem is provided by a spatially
periodic (lattice) potential~\cite{BBB,Ziad,BBB2,Ab05,Ze22}, or a trapping
harmonic-oscillator one~\cite{Al02,FNLS15}. A combination of self-focusing
cubic and defocusing quintic SPM terms in NLS equations offers another
possibility to suppress the collapse and stabilize solitons~\cite%
{Ga99,Zeng20}. Addition of higher-order dispersion to the usual second-order term may also
help to stabilize solitons and essentially modify their properties. In
particular, recent theoretical and experimental works have demonstrated how
temporal solitons may be maintained and shaped by the combination of the
second- and fourth-order GVDs \cite{Biancalana, two-GVD1,two-GVD2}. In a
similar context, it may be relevant to study possibilities for the
stabilization and control of solitons by a \emph{two-LI} setup, i.e., a
waveguide featuring a combination of two fractional-diffraction terms with
different values of LI, $\alpha _{1},\,\alpha _{2}\in (1,2]$ and the
corresponding real coefficients $a$ and $b$. In particular, a respective
generalization of the 1D version of one-LI FNLS equation (\ref{FNLS-0}) without a
potential is
\begin{equation}
i\frac{\partial \psi }{\partial z}=\frac{1}{2}\left[ a\left( -\frac{\partial
^{2}}{\partial x^{2}}\right) ^{\!\!\alpha _{1}/2}+b\left( -\frac{\partial
^{2}}{\partial x^{2}}\right) ^{\!\!\alpha _{2}/2}\right] \psi -g|\psi
|^{2}\psi .  \label{MUFNLS}
\end{equation}%
Note that the Kerr nonlinear term in Eq.~(\ref{MUFNLS}), $g|\psi |^{2}\psi $,
may be replaced by nonlinear terms of other types ($F(x, |\psi|^2)\psi$), such as the quintic one, $%
|\psi |^{4}\psi $, a generic term with the power-law nonlinearity, $|\psi
|^{2\sigma }\psi \, (\sigma>0) $, a combination of competing nonlinear terms, $g_{1}|\psi |^{2p}\psi
+g_{2}|\psi |^{2q}\psi\, (g_{1,2}\in\mathbb{R},\, p, q>0)$, logarithmic nonlinear term $\psi\ln|\psi|^2$, or by the saturable expression, $|\psi |^{2}\psi
/(1+S|\psi |^{2})\, (S>0)$~\cite{Ag00,Miha17,Kl14,Mal22,FNLS9,Xie21,Li22,Li23}.

A possibility to experimentally implement the two-LI system is actually
suggested by the above-mentioned work \cite{Li23}, where the fiber-laser
cavity included two holograms, one used for the emulation of the fractional
GVD, and the one forming the necessary shape of the input optical pulse. A
straightforward option is to use the two holograms to implement the action
of different fractional-GVD terms. In principle, structures emulating both
fractional terms can be inscribed on the same hologram, but using two
separate ones will facilitate the system's design.

We would like to develop the analysis of the two-LI fractional physical model  in the present work,
demonstrating that the additional fractional-diffraction term may indeed
help to suppress the critical collapse and stabilize solitons that would
otherwise be completely unstable. Further, formation of solitons is closely
related to the modulation instability (MI), alias the Benjamin--Feir
instability, which refers to the growth of perturbations on the
continuous-wave (CW) background~\cite{Be67,Za09}. The MI, caused by the
interplay of nonlinearity and dispersion, is a fundamentally important
phenomenon in various physical systems \cite%
{Ta86,Ag00,Ki15,Miha17,Bh21,Oa22,Mal22}. Spectra of the MI gain are
produced by means of the analysis of small modulational perturbations in the
framework of the linear approximation~\cite{Ag00,Ki00}. When the intensity
of the perturbation becomes comparable to that of the background CW, further
evolution of the MI is investigated by means of numerical simulations \cite%
{Tr91}. While the MI is usually studied in the framework of focusing
NLS equations, it also occurs in defocusing NLS media, a well-known example
being a system of cubic equations with the cross-phase-modulation
coefficient exceeding its SPM counterpart \cite{Ag00}. MI was also recently
explored~in the framework of FNLS equations \cite{Zh17,Zh22}. In the present
work, it is considered in the two-LI model (\ref{MUFNLS}).

Another fundamental phenomenon predicted by NLS equations is rogue waves
(RWs) that exist on top of the CW background subject to MI. An RW is an
isolated large-amplitude excitation that \textquotedblleft appears from
nowhere and disappears with no trace"~\cite{On13,Kh03,Ak10}. As a special
type of nonlinear waves, RWs have been found in nonlinear optics~\cite%
{So07,Ki10}, deep ocean~\cite{Ch11}, superfluids~\cite{superfluid}, plasmas,
\cite{RW_plasma}, BECs~\cite{bec-rw,bec-rw2}, atmosphere~\cite%
{Iafrati}, and even in financial markets~\cite{yanfrw}. Quantitative
relations between the MI and formation of RWs have been established in an
analytical form~\cite{Zh16,Li17}. From the viewpoint of MI, a resonant
perturbation on top of the CW background is a mechanism for the RW
generation.

To the best of our knowledge, RWs were not reported as solutions of
FNLS models before. In addition to the above-mentioned results for the solitons and
MI, the present paper reports the first- and second-order RWs existing on
the CW background in the two-LI FNLS equation. A noteworthy finding is that
first-order RWs can be generated in the case of the defocusing SPM term [and opposite signs
of the diffraction coefficients $a$ and $b$ in Eq. (\ref{MUFNLS})], which is
neither possible in the case of the integrable NLS equation with the regular
diffraction and cubic defocusing nonlinearity, nor as solutions of the
defocusing \textquotedblleft single-LI" FNLS equation, i.e., one with the
single diffraction term.

The main results reported in this paper are summarized as follows:

\begin{itemize}
\item {} We find families of fundamental-soliton solutions of the two-LI FNLS equation (\ref{MUFNLS}) by means of the
variational approximation (VA) and in a numerical form.

\item {} The additional fractional-diffraction term can be used to stabilize the family
of quasi-TS states existing in the FNLS model with only one LI $\alpha =1$.

\item {} The stability of solitons is explored in the two-LI FNLS equation by means of direct
simulations. The stability is also tested against collisions with outside one, two, or four
Gaussian pulses and adiabatic variations of the system parameters.

\item {} MI gain spectra are derived for the two-LI FNLS system under the consideration.
The two-LI FNLS equation with opposite signs of the diffraction coefficients
$a$ and $b$ admits MI even in the case of the defocusing nonlinearity. In
the absence of the MI, an analytically predicted boundary of the area
covered by propagating oscillatory perturbations is corroborated by
numerical results.

\item {} Applying excitations to the CW background, we obtain the first- and
second-order RWs of the two-LI FNLS equation. It is the first time when RWs are addressed in the
framework the fractional nonlinear-wave systems, to the best of our
knowledge.
\end{itemize}

The rest of this paper is arranged as follows. We introduce the stationary solutions of the two-LI FNLS
model (\ref{MUFNLS}) and some methods necessary to work with it in
Sec.~2. In Sec.~3, solitons are produced by dint of VA and numerical
methods. We specifically explore influence of the system parameters on
properties of the solitons. Robustness of the solitons against collisions
with impinging Gaussian pulses, and adiabatic evolution of solitons under
the action of slow variation of the system parameters are also reported.
In Sec.~4, we consider the occurrence of MI and first- and second-order RW
excitations in the two-LI FNLS equation with both focusing and defocusing
signs of the SPM term. The paper is concluded and discussed in Sec. 5.

%\section{The two-L\'{e}vy-index fractional model with the Riesz derivatives}

\section{Solitons of the two-LI FNLS equation: formation and stability}

While there are different definitions of fractional derivatives, the one of
the Riesz type in the two-LI FNLS Eq.~(\ref{MUFNLS}), which is actually realized in
fractional quantum mechanics {\color{red}\cite{La02}} and fractional optics \cite%
{Lo15,FNLS9}, is based on operators $\mathcal{F}$ and $\mathcal{F}^{-1}$ of
the direct and inverse Fourier transforms~\cite{La00,Lo15,Kl14,FNLS9}:
\begin{equation}
\begin{array}{rl}
\displaystyle\left( -\frac{\partial ^{2}}{\partial x^{2}}\right)
^{\!\!\alpha /2}\!\!\psi (x) & =\!\mathcal{F}^{-1}\left[ |p|^{\alpha }%
\mathcal{F}(\psi (x))\right] \vspace{0.1in} \\
& =\!\displaystyle\frac{1}{2\pi }\int_{-\infty }^{+\infty }\!dp|p|^{\alpha
}\!\!\int_{-\infty }^{+\infty }\!d\xi\,e^{ip(x-\xi)}\!\psi (\xi),\qquad
\alpha\in (1, 2],
\end{array}
\label{FT}
\end{equation}%
where $p$ is the wavenumber conjugate to the transverse coordinate $x$.
Other definitions of the fractional derivatives, such as Riemann-Liouville's
and Caputo's ones~\cite{VV13}, do not appear in the above-mentioned
realizations in quantum mechanics and optics. Equation~(\ref{MUFNLS}), in
which the fractional derivative is defined as per Eq. (\ref{FT}), can be
written in the variational form, $i\partial \psi /(\partial z)=\delta
\mathcal{H}/(\delta \psi ^{\ast })$, with Hamiltonian%
\begin{equation}
\mathcal{H}\!=\sum_{\alpha _{1},\alpha _{2}}\frac{\left\{ a,b\right\} }{4\pi
}\int_{-\infty }^{+\infty }dp|p|^{\alpha _{1,2}}\iint d\xi dxe^{ip(x-\xi
)}\psi ^{\ast }(x)\psi (\xi )\vspace{0.1in}-\frac{g}{2}\int_{-\infty
}^{+\infty }\left\vert \psi (x)\right\vert ^{4}dx,  \label{H}
\end{equation}%
where the definition (\ref{FT}) of the Riesz derivative is taken into
account, $\ast $ stands for the complex conjugate, and $\iint $ implies
integration over the plane of $\left( x,\xi \right) $.

Stationary solutions of Eq.~(\ref{MUFNLS}) are sought for as
%\begin{equation}
$\psi (x,z)=\phi (x;\mu )e^{-i\mu z},$  %\label{psiphi}
%\end{equation}%
where $-\mu $ is a real propagation constant, and a real amplitude function $%
\phi (x;\mu )$ obeys the stationary equation,
\begin{equation}
\mu \phi =\frac{1}{2}\left[ a\left( -\frac{\partial
^{2}}{\partial x^{2}}\right) ^{\!\!\alpha _{1}/2}+b\left( -\frac{\partial
^{2}}{\partial x^{2}}\right) ^{\!\!\alpha _{2}/2}\right] \phi
-g|\phi|^{2}\phi.  \label{ode21}
\end{equation}
which, via the direct-inverse Fourier transform (\ref{FT}), can also be rewritten as
\begin{equation}
\mu \phi =\frac{a}{4\pi }\int_{-\infty }^{+\infty }\!dp|p|^{\alpha
_{1}}\!\!\int_{-\infty }^{+\infty }\!d\xi e^{ip(x-\xi )}\!\phi (\xi )\!+\!%
\frac{b}{4\pi }\!\!\int_{-\infty }^{+\infty }\!dp|p|^{\alpha
_{2}}\!\!\int_{-\infty }^{+\infty }\!d\xi e^{ip(x\!-\!\xi )}\!\phi (\xi
)\!-\!g|\phi |^{2}\phi .\quad  \label{ode2}
\end{equation}%
It may be difficult to seek the analytical solutions of Eq.~(\ref{ode21}) or (\ref{ode2}).
Here we will used the modified squared-operator (MSO) method~\cite{Yang08}
to find numerical localized solitons of Eq.~(\ref{ode21}). To the end,
we firstly rewrite Eq.~(\ref{ode21}) as
\bee \label{Form}
  K_1\phi=(K_0-\mu)\phi=0,\quad
 K_{0}=\frac{1}{2}\left[ a\left(-\frac{\partial
^{2}}{\partial x^{2}}\right) ^{\!\!\alpha _{1}/2}+b\left(-\frac{\partial
^{2}}{\partial x^{2}}\right) ^{\!\!\alpha _{2}/2}\right] -g|\phi|^{2}.
  \ene
The stationary solutions $\phi(x)$ for the given propagation constant $\mu$ can be found by iterating as follows
\bee
\phi_{n+1}=\phi_n-\left(A^{-1}  {K}_2  {A}^{-1}  {K}_1 \phi_n-c_n\left\langle {B}_n,\,  {K}_1  {A}^{-1}  {K}_1 \phi\right\rangle  {B}_n\right)\Delta x,
\ene
where
\bee\no
c_n=\frac{1}{\left\langle {AB}_n,  {B}_n\right\rangle}-\frac{1}{\left\langle {K}_2 B_n,  A^{-1}  K_2  B_n\right\rangle \Delta x},
\quad
 B_n=\phi_n-\phi_{n-1},
\ene
with $<\cdot,\cdot>$ denoting the inner product in $L^2$ space, $A$ a real-valued positive-definite and Hermitian acceleration operator, and $K_2$ the linearization operator of Eq.~(\ref{ode21})
\bee
K_2=\frac{1}{2}\left[ a\left(-\frac{\partial
^{2}}{\partial x^{2}}\right) ^{\!\!\alpha _{1}/2}+b\left(-\frac{\partial
^{2}}{\partial x^{2}}\right) ^{\!\!\alpha _{2}/2}\right]-3g \phi^2-\mu.
\ene

Once stationary solutions to Eq.~(\ref{ode21}) or Eq.~(\ref{ode2}) were found, their stability can be
explored by means of direct simulations of Eq. (\ref{MUFNLS}) with the
perturbed input, $\psi (x,0)=\phi (x)(1+\epsilon )$, where $\epsilon $ is a
random perturbation, whose amplitude is taken at a $2\%$ level. In our
numerical simulations, the spatial differention was carried out by means of
the discrete Fourier transform, and we adopt an explicit fourth-order
Runge-Kutta scheme to advance along propagation distance $z$~\cite{Yang08}.

\subsection{The variational approximation (VA) and dynamics}

Localized solutions of the two-LI FNLS equation (\ref{ode2}) can be sought for in an
approximate analytical form in the framework of VA~\cite%
{FNLS9,Qiu20,Ze22,Sakaguchi}. For this purpose, Eq.~(\ref{ode2}) can be
derived from the Lagrangian, cf. expression (\ref{H}) for the Hamiltonian:%
\begin{eqnarray}
{\cal L} &=&-\frac{\mu }{2}\int_{-\infty }^{+\infty }dx\phi ^{2}(x)+\frac{a}{8\pi }%
\int_{-\infty }^{+\infty }dp|p|^{\alpha _{1}}\iint d\xi dxe^{ip(x-\xi )}\phi
(x)\phi (\xi )\vspace{0.1in}  \notag \\
&&\qquad +\frac{b}{8\pi }\int_{-\infty }^{+\infty }dp|p|^{\alpha _{2}}\iint
d\xi dxe^{ip(x-\xi )}\phi (x)\phi (\xi )-\frac{g}{4}\int_{-\infty }^{+\infty
}dx\phi ^{4}(x).  \label{LL}
\end{eqnarray}%

A simple form of the variational ansatz approximating solitons of Eq.~(\ref%
{ode2}) is provided by the Gaussian,
\begin{equation}
\phi (x)=A\exp [-x^{2}/(2W^{2})]  \label{Gauss}
\end{equation}%
with real-valued parameters $A\,$and $W$ representing the amplitude and
width, respectively. The power (\ref{P}) of ansatz (\ref{Gauss}) is
\begin{equation}
P_{A}\!=\!\!\sqrt{\pi }A^{2}W.\qquad  \label{PA}
\end{equation}
The substitution of the Gaussian ansatz in Lagrangian (\ref{LL}) yields the
corresponding effective (spatially integrated) Lagrangian:
%\begin{widetext}
\begin{equation}
{\cal L}_{\mathrm{eff}}=\displaystyle-\frac{\sqrt{\pi }}{2}\mu A^{2}W+\frac{a}{4}%
\Gamma \left( \frac{\alpha _{1}+1}{2}\right) A^{2}W^{1-\alpha _{1}}+\frac{b}{%
4}\Gamma \left( \frac{\alpha _{2}+1}{2}\right) A^{2}W^{1-\alpha _{2}}-\frac{g%
}{4}\sqrt{\frac{\pi }{2}}A^{4}W.
\end{equation}%
Notice that one may use other types of soliton ansatze (e.g., $A{\rm sech}(Wx)$), not the Gaussian (\ref{Gauss}), such that
the corresponding effective (spatially integrated) Lagrangian may not be explicitly given, but can be found
by using numerical integral methods.

\v It is more convenient to write it in terms of $W$ and power (\ref{PA}),
eliminating amplitude wth the help of Eq. (\ref{PA}) as%
\begin{equation}
{\cal L}_{\mathrm{eff}}=\!\!\displaystyle-\frac{\mu }{2}P_{A}+\frac{a}{4\sqrt{\pi }}%
\Gamma \left( \frac{\alpha _{1}+1}{2}\right) P_{A}W^{-\alpha _{1}}+\frac{b}{4%
\sqrt{\pi }}\Gamma \left( \frac{\alpha _{2}+1}{2}\right) P_{A}W^{-\alpha
_{2}}-\frac{g}{4\sqrt{2\pi }}P_{A}^{2}W^{-1}.  \label{Leff}
\end{equation}%
Then, values of $P_{A}$ and $W$ are predicted by the Euler-Lagrange
equations, $\partial {\cal L}_{\mathrm{eff}}/\partial P_{A}=\partial {\cal L}_{\mathrm{eff}%
}/\partial W=0$, that is,%
\begin{eqnarray}
2\sqrt{\pi }\mu -a\Gamma \left( \frac{\alpha _{1}+1}{2}\right) W^{-\alpha
_{1}}-b\Gamma \left( \frac{\alpha _{2}+1}{2}\right) W^{-\alpha _{2}}+\sqrt{2}%
gP_{A}/W &=&0,\vspace{0.15in}  \label{VA} \\
4\sqrt{\pi }\mu +2a(\alpha _{1}-1)\Gamma \left( \frac{\alpha _{1}+1}{2}%
\right) W^{-\alpha _{1}}+2b(\alpha _{2}-1)\Gamma \left( \frac{\alpha _{2}+1}{%
2}\right) W^{-\alpha _{2}}+\sqrt{2}gP_{A}/W &=&0.  \label{VA2}
\end{eqnarray}%
Eliminating $P_{A}$ from Eqs. (\ref{VA}) and (\ref{VA2}), one arrives at a
single VA-predicted equation for the width, %\bee
%\begin{array}{l}
%\d\sqrt{\pi}\mu+a\left(\alpha_1-\frac12\right)\Gamma\left(\frac{\alpha_1+1}{2}\right)W^{-\alpha_1}\v\\
%\quad \d +b\left(\alpha_2-\frac12\right)\Gamma\left(\frac{\alpha_2+1}{2}\right)W^{-\alpha_2}=0,
%\end{array}
%\ene
\begin{equation}
\begin{array}{l}
\displaystyle\sqrt{\pi }\mu +a\left( \alpha _{1}-\frac{1}{2}\right) \Gamma
\left( \frac{\alpha _{1}+1}{2}\right) W^{-\alpha _{1}}+b\left( \alpha _{2}-%
\frac{1}{2}\right) \Gamma \left( \frac{\alpha _{2}+1}{2}\right) W^{-\alpha
_{2}}=0,%
\end{array}
\label{W}
\end{equation}%
which can then be solved numerically.

\begin{figure}[t]
\centering
\vspace{-0.15in} {\scalebox{0.75}[0.75]{\includegraphics{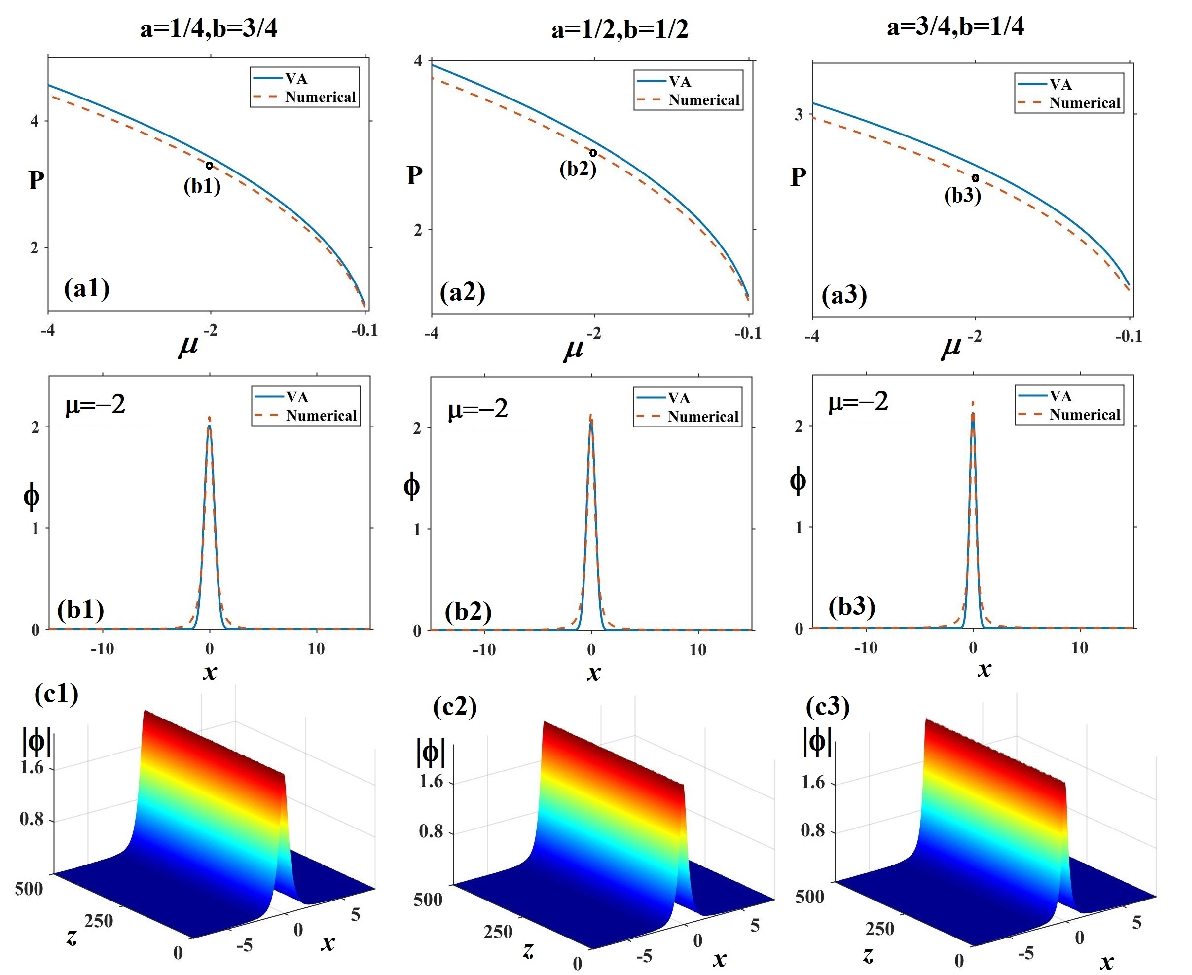}}}\hspace{-0.3in}
\vspace{0.1in}
\caption{{\protect\small The dependence of the soliton's power $P$, as
produced by the VA and numerical solutions of Eq.~(\protect\ref{ode2}) with $%
\protect\alpha _{1}=1,\,\protect\alpha _{2}=1.8,\,$and $g=1$, on propagation
constant $-\protect\mu $ at (a1) $a=1/4,\,b=3/4$; (a2) $a=b=1/2$, and (a3) $%
a=3/4,\,b=1/4$. The labels in panels (a1,a2,a3) correspond to the solitons
shown in (b1,b2,b3), respectively. Panels (b1-b3) display the corresponding
profiles of stable solitons predicted by the VA and produced by the
numerical solution (solid and dashed lines, respectively) for $\protect\mu %
=-2$. (c1-c3) Simulations of the perturbed evolution of the solitons from
panels (b1-b3), corroborating their stability$.$ }}
\label{P1}
\end{figure}

The first noteworthy finding provided by the VA and numerical results alike
is that while, in the above-mentioned critical case of $\alpha _{1}=1$, Eq. (%
\ref{ode2}) with the self-focusing sign of the nonlinearity, $g=+1$,
produces a family of quasi-TS solutions, with the single value of the power
(norm),
\begin{equation}
\left( P_{\text{Townes }}\right) _{\mathrm{VA}}=\sqrt{2}  \label{PTownes}
\end{equation}%
[its numerically found counterpart is given above by Eq. (\ref{Pcrit})] \cite%
{Qiu20}, the addition of the diffraction term with another value of LI, $%
\alpha _{2}$, to Eq. (\ref{ode2}) lifts the norm degeneracy, as shown in
Fig.~\ref{P1}. The figure also demonstrates high accuracy provided by the VA
for the resulting dependences $P_{A}(\mu )$ and typical shapes of individual
solitons. A small discrepancy in the shape of the tails of the solitons is
explained by the fact that the inherent scaling of Eq. (\ref{ode2}) implies
that true localized solutions have an asymptotic form $\phi _{\alpha
}(x)\sim |x|^{-1-\alpha }$ at $|x|\rightarrow \infty $,
where $\alpha $ is the smaller value from $\alpha _{1,2}$ \cite{Kl14}, while
the Gaussian ansatz (\ref{Gauss}) does not include such tails.

\begin{figure}[t]
\centering
\vspace{-0.15in} {\scalebox{0.82}[0.82]{\includegraphics{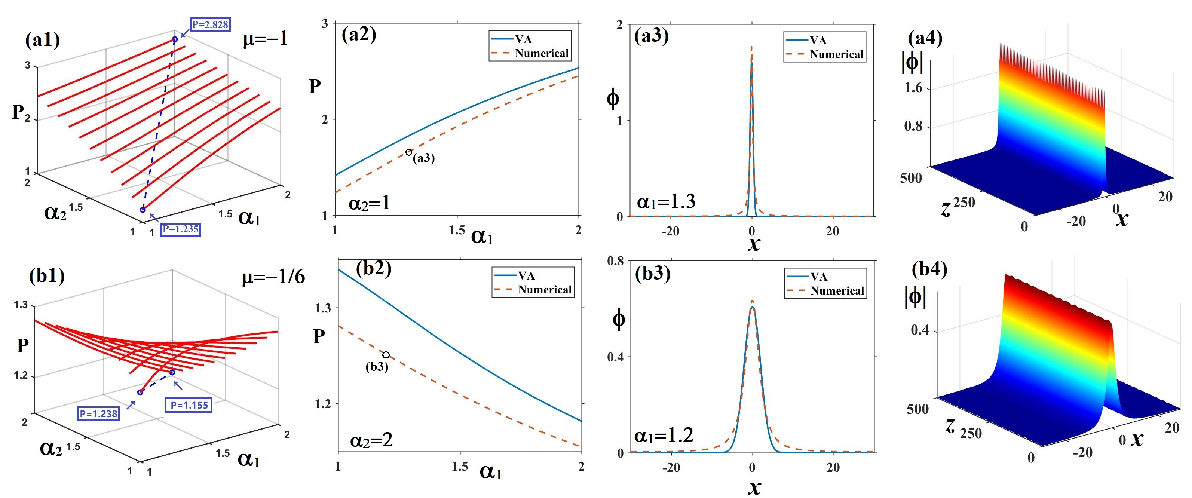}}}\hspace{-0.3in}
\vspace{0.1in}
\caption{{\protect\small Panels (a1) and (b1): the dependence of the
soliton's power $P$, as produced by the VA and numerical solutions of Eq.~(\protect\ref{ode2})
on LIs $(\protect\alpha _{1},%
\protect\alpha _{2}),$ for a fixed propagation constant $\protect\mu =-1$
(a1) and $\protect\mu =-1/6$ (b1). Panel (a2): a cross section of (a1) drawn
at $\protect\alpha _{2}=1$. (a3,a4): The shape and perturbed evolution of a
weakly unstable soliton for $\protect\alpha _{1}=1.3$, which
corresponds to the point marked in (a2). Panel (b2): a cross
section of (b1) drawn at $\protect\alpha _{2}=2$. (b3,b4): The shape and
perturbed evolution of a stable soliton for $\protect\alpha %
_{1}=1.2$, which corresponds to the point marked in (b2). Other parameters
are $a=1/2,\,b=1/2,\,g=1.$ }}
\label{P2}
\end{figure}

Note that the $P_{A}(\mu )$ dependences displayed in Figs.~\ref{P1}(a1-a3)
satisfy the Vakhitov-Kolokolov criterion, $dP/d\mu <0$, which is the
well-known necessary stability condition for solitons supported by a
self-focusing nonlinearity, which, however, is not sufficient for the
stability in all cases \cite{VK,Be98}. Systematic numerical simulations
indicate that the soliton families presented in Figs.~\ref{P1}(a1-a3) are
completely stable in the interval of $\mu \in \lbrack -4,0]$. In particular,
the stability of the solitons presented in Figs. \ref{P1}(b1-b3) is directly
corroborated by the simulations of their perturbed evolution, as shown in
Figs. \ref{P1}(c1-c3), respectively. Depending on valaues of the parameters,
stable solitons may also be found at $\mu <-4$, but we here do not aim to
consider the extension to that region in detail.

The arrest of the critical collapse in the two-LI model can be explained
similar to the consideration of the onset of the collapse in the case of the
regular diffraction ($\alpha =2$)~\cite{Za12,Mal22,Fi15}. To this end, the
Hamiltonian for Eq.~(\ref{MUFNLS}) can be divided in two parts, which
represent, respectively, the two diffraction terms and the self-focusing
one. Considering a localized state with radius $L$ and amplitude $B$, an
obvious estimate of the power is $P\sim B^{2}L.$ Similarity, the diffraction
and self-focusing cubic terms in the Hamiltonian Eq.~(\ref{H}) can be
estimated as:
\begin{equation}
{\cal H}_{\text{diff}}\sim aP/L^{\alpha _{1}}+bP/L^{\alpha _{2}},\qquad {\cal H}_{\text{%
focusing}}\sim -gP^{2}/L.
\end{equation}%
The critical collapse occurs if $|{\cal H}_{\text{focusing}}|$ and ${\cal H}_{\text{diff}}$
scale as the same negative power of $L$ at $L\rightarrow 0$ for fixed $P$
(recall that $P$ is the dynamical invariant). Consequently, if only the
single-fractional-diffraction term is considered, then $\alpha _{1}=1$
corresponds to the critical collapse. However, with the introduction of the
second diffraction term, $\alpha _{2}>1$, one may expect that the critical
collapse may be arrested in the case of $\alpha _{1}=1$, as confirmed by the
present numerical results.

Dependences of the soliton's power on the two LI values, $\alpha _{1}$ and $%
\alpha _{2}$, while the corresponding diffraction coefficients are fixed as $%
a=b=1/2$, are shown in Figs. \ref{P2}(a1,b1,a2,b2), for propagation
constants $-\mu =1$ and $1/6$ in panels (a1) and (b1), respectively. In this
connection, it is relevant to mention that, in the limit case of $\alpha
_{1}=\alpha _{2}=1$, the solitons form the TS family, with the unique value
of the power predicted by the VA,
\begin{equation}
\left( P_{\text{Townes }}\right) _{\mathrm{VA}}^{\left( a,b\right) }=\sqrt{2}%
(a+b)>0,  \label{Tw}
\end{equation}%
cf. Eq. (\ref{PTownes}). On the other hand, in the case of the regular
(non-fractional) diffraction, $\alpha _{1}=\alpha _{2}$, the usual NLS
solitons have the power
\begin{equation}
P_{\alpha _{1}=\alpha _{2}=2}=2\sqrt{-2\mu (a+b)}.  \label{ES}
\end{equation}

Comparing the values of $P$ given by Eqs.~(\ref{ES}) and (\ref{Tw}), one can
conclude that the $P(\alpha _{1},\alpha _{2})$ should be a decaying function
of LIs in the case of $0<-\mu <(a+b)/4$. This conjecture is corroborated by
the dependences $P(\alpha _{1},\alpha _{2})$ shown in Figs. \ref{P2}(b1,b2).
Note that the accuracy of the VA for $\mu =-1/6$ is poorer in Fig. \ref{P2}%
(b2) than in Fig. \ref{P2}(a2) for $\mu =-1$. This is explained by the fact
that the soliton's shape is flatter for the essentially smaller value of $%
|\mu |$, which is poorer approximated by the Gaussian ansatz (\ref{Gauss}).

As concerns the stability, the numerical investigation reveals that the
soliton branch shown in Fig. \ref{P2}(b2) for $\mu =-1/6$ is completely
stable, see an example presented in Fig. \ref{P2}(b4). The result is
different for the branch with $\mu =-1$, which is shown in Fig. \ref{P2}%
(a2): it is stable at $\alpha _{1}\geq 1.3$, and unstable at $\alpha
_{1}<1.3 $ [the (in)stability was verified with interval $\Delta \alpha
_{1}=0.1$]. An example of the (weak) instability is presented in Fig. \ref%
{P2}(a4), in the from of small-amplitude oscillations setting in on top of
the soliton.

\begin{figure}[t]
\centering
\vspace{-0.15in} {\scalebox{0.75}[0.75]{\includegraphics{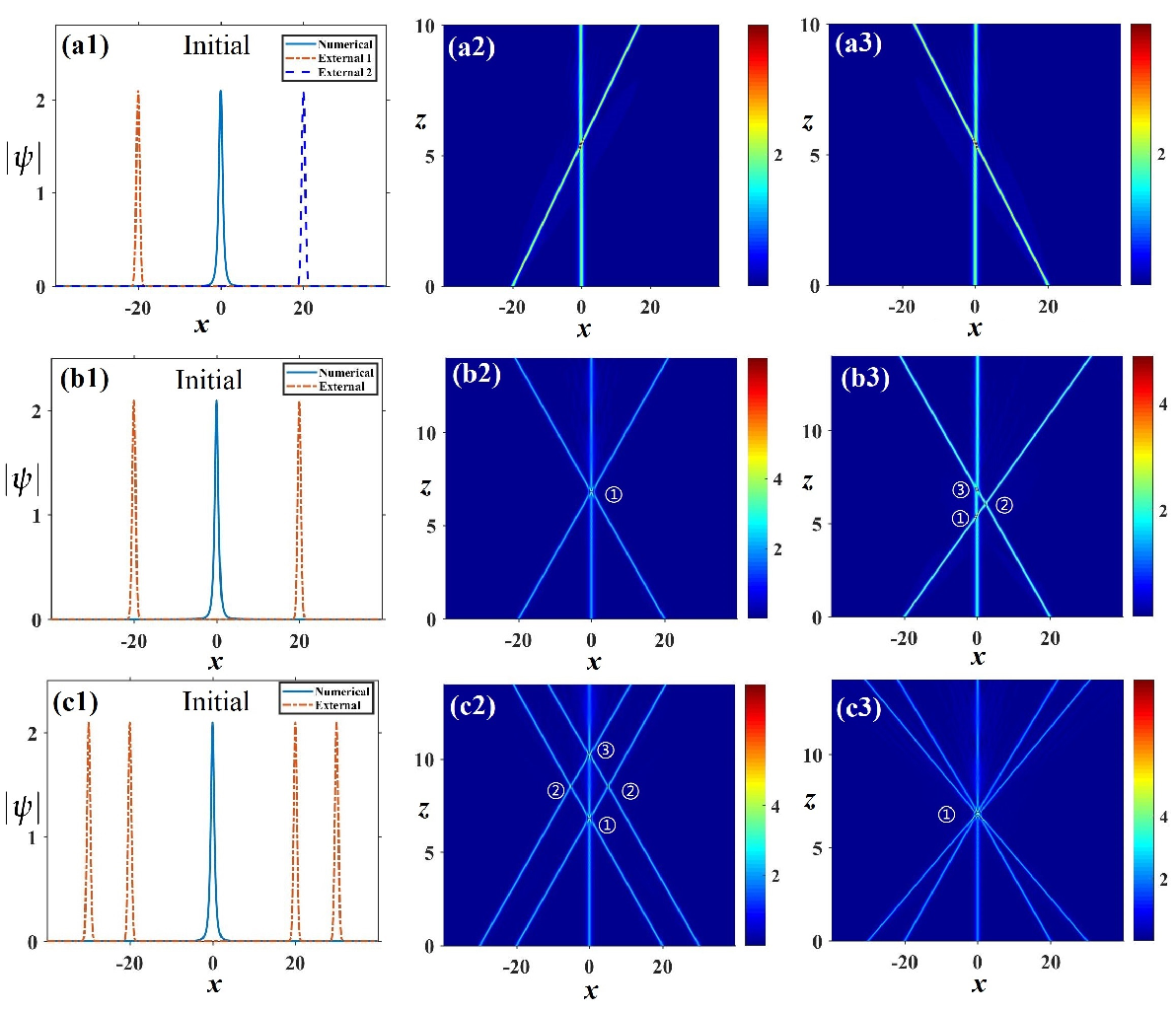}}}\hspace{-0.3in%
} \vspace{0.1in}
\caption{{\protect\small (a1) The distribution of }$|\protect\psi (x)|$%
{\protect\small \ in the initial conditions represented by Eq.~(\protect\ref%
{In0}). (a2,a3): The interaction pictures in these cases. (b1) The
distribution of }$|\protect\psi (x)|${\protect\small \ in the initial
conditions~represented by Eqs. (\protect\ref{In1}) and (\protect\ref{In2}).
(b2) and (b3): The interaction pictures in these cases. (c1) The distribution
of }$|\protect\psi (x)|${\protect\small \ in the initial conditions
represented by Eqs.~ (\protect\ref{In3}) and (\protect\ref{In4}). (c2) and
(c3): The interaction pictures in these cases. Encircled digits are numbers
of consecutive collisions. Other parameters are }$\protect\alpha _{1}=1.0$, $%
\protect\alpha _{2}=1.8$, $a=1/4$, $b=3/4$, $g=1.${\protect\small \ }}
\label{Inter}
\end{figure}

\subsection{Elastic collisions of solitons with impinging pulses}

To additionally explore robustness of the solitons in the two-LI FNLS model (\ref{MUFNLS}), it is also relevant to
check if they keep their integrity against \textquotedblleft bombardment" by
impinging pulses. In this connection, we note that the fractional
diffraction destroys the Galilean invariance of Eq. (\ref{MUFNLS}), but the
application of a large kick to a quiescent localized state sets it in motion
in approximately the same fashion as it happens in the Galilean-invariant
NLS equations with regular diffraction \cite{FNLS9}. To implement the
robustness analysis, we examine collisions between the soliton, $\phi (x)$,
displayed in Fig.~\ref{P1}(b1) with kicked Gaussian pulses. The following
characteristic examples demonstrate the robustness of the solitons against
the collisions.

\textit{Case 1.}---We start with the collision between the soliton
and a single moving Gaussian, generated by the input:
\begin{equation}
\psi ^{\pm }(x,z=0)=\phi (x)+2.094e^{-(x\pm x_{0})^{2}/(2w^{2})\pm ivx}%
\vspace{0.1in},  \label{In0}
\end{equation}%
where the chosen amplitude $2.094$ coincides with the amplitude of the
stationary solution $\phi (x)$, so the two amplitudes are equal, see Fig.~%
\ref{Inter}(a1), while $x_{0},w{\ }$and $v$ are the center and width of the
Gaussian, and the kick applied to it. We set $x_{0}=20,w=0.4$ and $v=6$
here. The collisions are completely elastic, as shown in Figs.~\ref{Inter}%
(a2) and (a3), respectively. The identical results corresponding to $\pm $
in Eq. (\ref{In0}) confirm the absence of an instability which could lead to
spontaneous symmetry breaking.

\vspace{0.1in} \textit{Case 2.}---We also consider the collision
with a pair of identical Gaussians impinging upon the soliton from opposite
directions, initiated by
\begin{equation}
\begin{array}{rl}
\psi (x,0)= & \phi
(x)+2.094e^{-(x-x_{0})^{2}/(2w^{2})-ivx}+2.094e^{-(x+x_{0})^{2}/(2w^{2})+ivx},%
\end{array}
\label{In1}
\end{equation}%
where we set $x_{0}=20,w=0.4$, and $v=6$. This input is exhibited in Fig.~%
\ref{Inter}(b1). It is seen in Fig. \ref{Inter}(b2) that the collision is
fully elastic in this case too.

\vspace{0.1in}\textit{Case 3.}---The collision with the Gaussians
impinging with different velocities. In this case, the initial condition is
chosen as
\begin{equation}
\begin{array}{rl}
\psi (x,0)= & \phi
(x)+2.094e^{-(x-x_{0})^{2}/(2w^{2})-iv_{1}x}+2.094e^{-(x+x_{0})^{2}/(2w^{2})+iv_{2}x},%
\end{array}
\label{In2}
\end{equation}%
where $x_{0}=20,w=0.4$, and $v_{1}=6,v_{2}=8$. The distribution of $|\psi
(x)|$ in this input is the same as in Fig.~\ref{Inter}(c1). The asymmetry of
the input gives rise to  two separate collisions in Fig.~\ref{Inter}(b3),
the outcome remaining completely elastic.

\vspace{0.1in}\textit{Case 4.}---A double collision with two pairs
of counterpropagating Gaussians, generated by input
\begin{equation}
\begin{array}{rl}
\label{In3}\psi (x,0)= & \phi
(x)+2.094e^{-(x-x_{0})^{2}/(2w^{2})-ivx}+2.094e^{-(x+x_{0})^{2}/(2w^{2})+ivx}%
\vspace{0.1in} \\
& \qquad +2.094e^{-(x-x_{1})^{2}/(2w^{2})-ivx}+2.094e^{-(x+x_{1})^{2}/(2w^{2})+ivx},%
\end{array}%
\end{equation}%
where $x_{0}=20,x_{1}=30,w=0.4,v=6$, as shown in Fig.~\ref{Inter}(c1). In
this case, the impinging Gaussians also collide between themselves. In spite
of the more complex arrangement, the soliton is not disturbed by the
collisions, as seen in Fig.~\ref{Inter}(c2).

\vspace{0.1in} \textit{Case 5.}---Lastly, an example of the
strongest simultaneous collision of the solitons with four impinging
Gaussians, the respective input being
\begin{equation}
\begin{array}{rl}
\label{In4}\psi (x,0)= & \,\phi
(x)+2.094e^{-(x-x_{0})^{2}/2w^{2}-iv_{0}x}+2.094e^{-(x+x_{0})^{2}/2w^{2}+iv_{0}x}%
\vspace{0.1in} \\
& \qquad
+2.094e^{-(x-x_{1})^{2}/2w^{2}-iv_{1}x}+2.094e^{-(x+x_{1})^{2}/2w^{2}+iv_{1}x},%
\end{array}%
\end{equation}%
where $x_{0}=20,x_{1}=30,w=0.4,v_{1}=6,v_{2}=10$. The distribution of $|\psi
(x)|$ in this input is the same as in Fig.~\ref{Inter}(c1). It is seen in
Fig.~\ref{Inter}(c3) that the strongest collision does not produce any
destabilization of the soliton either.

\begin{figure}[t]
\centering
\vspace{-0.15in} {\scalebox{0.7}[0.7]{\includegraphics{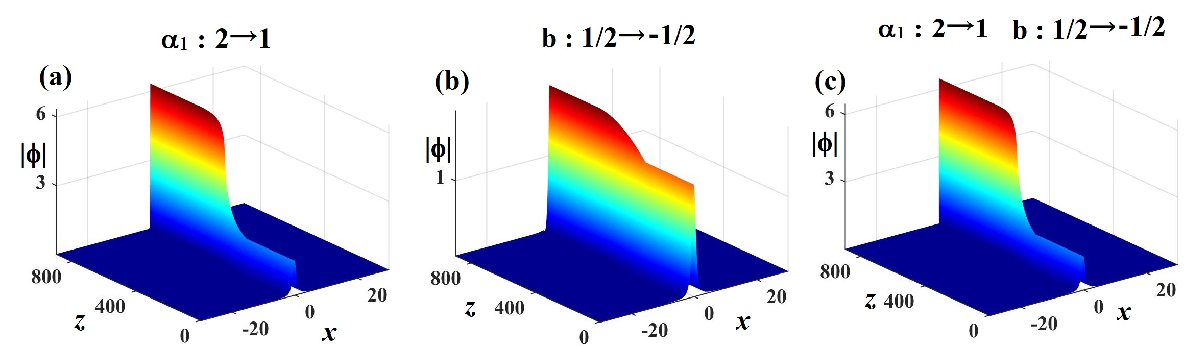}}}\hspace{-0.3in}
\vspace{0.1in}
\caption{{\protect\small The application of the adiabatic transformation (%
\protect\ref{condition}) to the model based on Eq. (\protect\ref{MUFNLS}).
(a) The variation of LI $\protect\alpha _{1}$. (b) The variation of the
diffraction coefficient $b$. (c) The simultaneous variation of $\protect%
\alpha _{1}$ and $b$. Other parameters are $a=1/2,\protect\alpha _{2}=1.0$
and $g=1$. }}
\label{Ex}
\end{figure}

\subsection{Soliton dynamics under the action of adiabatic modulations}

Here we aim to study evolution of stable solitons driven by adiabatic
variations of the system parameters. We focus on the modulation format in
which the LI of the first fractional-diffraction term and the coefficient in
front of the second one are made slowly varying functions of the propagation
distance:$\ \alpha _{1}\rightarrow \alpha _{1}(z)$ and $b\rightarrow b(z)$,
cf. Refs. \cite{FNLS21,yan15}. Accordingly, Eq.~(\ref{MUFNLS}) is replaced
by the following generalized one:
\begin{equation}
i\frac{\partial \psi }{\partial z}=\frac12\left[ a\left( -\frac{\partial ^{2}}{%
\partial x^{2}}\right) ^{\!\!\alpha _{1}(z)/2}+b(z)\left( -\frac{\partial
^{2}}{\partial x^{2}}\right) ^{\!\!\alpha _{2}/2}\right] \psi -g|\psi
|^{2}\psi ,  \label{MFNLS2}
\end{equation}%
where the $z$-dependent parameters $\left\{ \alpha _{1}(z),b(z)\right\} $
are taken as the switch function
\begin{equation}
\left\{ \alpha _{1}(z),b(z)\right\} \!=\!\left\{ \!\begin{aligned} & \left\{
\alpha _{1i},b_{i}\right\}, & 0\leq z \leq 300,\\ &\left\{ \alpha
_{1i},b_{i}\right\} \!+\!\left\{ \left( \alpha _{1e}-\alpha _{1i}\right)
,\left( b_{e}-b_{i}\right) \right\} \!\sin \left[ \frac{(z\!-\!300)}{600}\pi
\right], & 300< z \leq 600, \\ &\left\{ \alpha _{1e},b_{e}\right\}, & 600<
z\leq z_{\text{max}} \end{aligned}\right.  \label{condition}
\end{equation}%
with subscripts $i$ and $e$ referring to the initial and final (eventual)
values in the parameters, at $z=0$ and $z=z_{\text{max}}$, respectively.
Numerical results are presented here in Fig. \ref{Ex} for the initial state
corresponding to the stationary solution with $a=b=1/2,\,\alpha
_{1}=2.0,\,\alpha _{2}=1.0$ and $\mu =-1$. The results may be summarized as
follows:

\begin{itemize}
\item {} The adiabatic variation of LI $\alpha _{1}(z)$ as per Eq. (\ref%
{condition}), while $b=1/2$ is kept constant. In this case, the initial
stable localized mode with parameters $\left( \alpha _{1i},b\right)
=(2.0,0.5)$ is smoothly transformed into another stable mode, corresponding
to parameters $\left( \alpha _{1e},b\right) =(1.0,0.5)$. It is seen from
Fig.~\ref{Ex}(a) that the soliton undergoes three stages of the evolution:
at first, stable propagation of the initial state; then, the gradual
increase of the amplitude from $1.496$ to $6.345$; eventually, stable
propagation of the final state.

\item {} The adiabatic variation of the diffraction coefficient $b$, while $%
\alpha _{1}=2.0$ is kept constant. In this case, it is seen in Fig.~\ref{Ex}%
(b) that the stable initial mode with parameters $\left( \alpha
_{1},b_{i}\right) =(2.0,0.5)$ is smoothly transformed into another stable
mode, corresponding to $\left( \alpha _{1},b_{e}\right) =(2.0,-0.5)$ (the
one with opposite signs of the diffraction coefficients $a$ and $b$).
Similar to the previous case, the transition leads to an increase of the
soliton's amplitude.

\item {} The simultaneous variation of both LI $\alpha _{1}$ and diffraction
coefficient $b$. The result is a gradual transition from the stable mode
corresponding to $\left( \alpha _{1i},b_{i}\right) =(2.0,0.5)$ to one with $%
\left( \alpha _{1e},b_{e}\right) =(1.0,-0.5)$. In this case too, a strong
increase of the soliton's amplitude is a result of the adiabatic
transformation.
\end{itemize}

\section{Modulational instability of the CW state and rogue-wave excitations}

\subsection{The modulational instability}

The CW solution of Eq.~(\ref{MUFNLS}), with power $P$, is
\begin{equation}
\psi _{\mathrm{CW}}(x,z)=\sqrt{P}\exp {(igPz)}.  \label{CW}
\end{equation}%
To address the modulational instability (MI) of this state, we set $P=1$ by means of scaling. The
standard linear-stability analysis is performed by adding a small
perturbation to the CW solution,
\begin{equation}
\psi (x,z)=\left[ 1+\Psi(x,z)\right] e^{igz}  \label{Ln}
\end{equation}%
with the small perturbation subject to the usual constraint, $|\Psi(x,z)|\ll 1$.
Substituting ansatz (\ref{Ln}) into Eq.~(\ref{MUFNLS}) and linearizing the
result with respect to $\Psi(x,z)$ leads to the following evolution equation for the
perturbation:
\begin{equation}
i\frac{\partial \Psi}{\partial z}-\frac12\left[ a\left( -\frac{\partial ^{2}}{\partial
x^{2}}\right) ^{\alpha _{1}/2}+b\left( -\frac{\partial ^{2}}{%
\partial x^{2}}\right) ^{\alpha_{2}/2}\right] \Psi+g(\Psi+\Psi^{\ast
})=0.  \label{MLE}
\end{equation}%
The solution to Eq. (\ref{MLE}) with wavenumber $k$ and propagation constant
$w$ is looked for as
\begin{equation}
\Psi(x,z)=f_{1}e^{i(kx-w(k)z)}+f_{2}e^{-i(kx-w(k)z)},  \label{Mo}
\end{equation}%
where $f_{1,2}$ are constant amplitudes. Substituting expression~(\ref{Mo})
in Eq.~(\ref{MLE}), one obtains the dispersion relation
\begin{equation}
w^{2}(k)=\frac{1}{4}\left( a|k|^{\alpha _{1}}+b|k|^{\alpha _{2}}\right)
^{2}-g\left( a|k|^{\alpha _{1}}+b|k|^{\alpha _{2}}\right) .\quad
\label{relation}
\end{equation}

According to the linear-stability theory, MI occurs for $w^{2}(k)<0$. In
this case, the instability growth is found as%
\begin{equation}
G=\frac{1}{2}\sqrt{\left( a|k|^{\alpha _{1}}+b|k|^{\alpha _{2}}\right) \left[
4g-\left( a|k|^{\alpha _{1}}+b|k|^{\alpha _{2}}\right) \right] }.\quad
\label{G}
\end{equation}%
If the MI takes place, i.e., Eq. (\ref{G}) has a positive expression under
the square root, the perturbations exponentially grow with rate (\ref{G}) in
intervals of $k$ determined by conditions
\begin{equation}
0<a|k|^{\alpha _{1}}+b|k|^{\alpha _{2}}<4g,\quad \text{for}\quad g>0,
\label{g>0}
\end{equation}%
\begin{equation}
4g<a|k|^{\alpha _{1}}+b|k|^{\alpha _{2}}<0,\quad \text{for}\quad g<0,
\label{g<0}
\end{equation}%
with the largest gain, $G_{\text{max}}=|g|$ attained at $a|k|^{\alpha
_{1}}+b|k|^{\alpha _{2}}=2g$. Note that LIs $\alpha _{1,2}\,\ $determine the
bandwidth of the gain but they do not affect the value of $G_{\text{max}}$.

\begin{figure}[!t]
\centering
\vspace{-0.15in} {\scalebox{0.8}[0.8]{\includegraphics{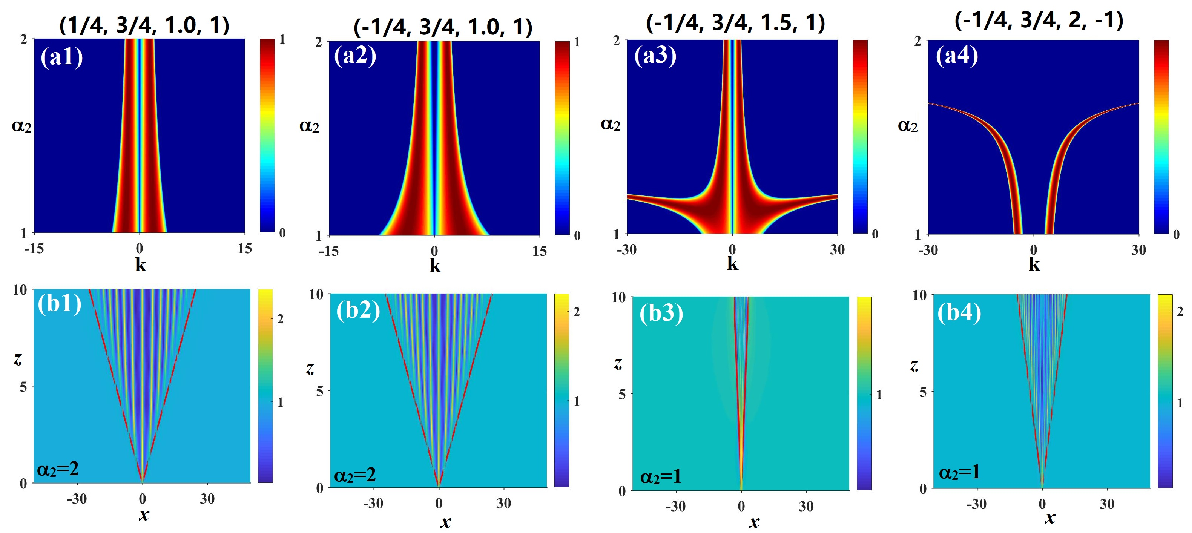}}}\hspace{-0.3in}
\vspace{0.1in}
\caption{{\protect\small (a1-a4): The MI gain in the $(k,\protect\alpha %
_{2}) $ plane with different parameters. (a1): $(a,b,\protect\alpha %
_{1},g)=(1/4,3/4,1.0,1)$; (a2): $(a,b,\protect\alpha %
_{1},g)=(-1/4,3/4,1.0,1) $; (a3): $(a,b,\protect\alpha %
_{1},g)=(-1/4,3/4,1.5,1)$; (a4): $(a,b,\protect\alpha %
_{1},g)=(-1/4,3/4,2,-1) $. (b1-b4): The propagation initiated by inputs
given by Eq.~(\protect\ref{Ini}). The red lines denote the boundary
predicted as $x=\pm w^{\prime }(k)_{\text{min}}t$, see Eq. (\protect\ref%
{boundary}).}}
\label{MI}
\end{figure}

We then consider the MI gain for different sets of the system parameters:

\begin{itemize}
\item {} First, set $a=1/4,b=3/4,\alpha _{1}=1,g=1$. The corresponding
instability region in the $(k,\alpha _{2})$ plane is depicted in Fig.~\ref%
{MI}(a1). It follows from Eq. (\ref{g>0}) that the MI take place in the
interval of
\begin{equation}
0<|k|+3|k|^{\alpha _{2}}<16,  \label{MI1}
\end{equation}%
with the maximum gain attained at $|k|+3|k|^{\alpha _{2}}=8$. It is seen
from Fig.~\ref{MI}(a1) that the MI\ bandwidth is quite sensitive to values $%
\alpha _{2}$, the bandwidth decreasing when $\alpha _{2}$ varies from $1$ to
$2$. In this case, the bandwidth reduction is a direct corollary of Eq. (\ref%
{MI1}). We also explore boundaries of the area occupied by oscillatory
perturbations which propagate on top of the underlying CW in the case when
the MI does not occur, for $\alpha _{2}=2$. The boundaries are established
by trajectories of the slowest waves generated by the initial perturbation,
which, in turn, are determined by the minimum of the group velocity $%
w^{\prime }(k)=dw(k)/dk$. For $k>0$, the group velocity is
obtained from Eq. (\ref{relation}) as
\begin{equation}
w^{\prime }(k)\!=\!\frac{\left( a\alpha _{1}k^{\alpha _{1}-1}\!+\!b\alpha
_{2}k^{\alpha _{2}-1}\right) \!(ak^{\alpha _{1}}\!+\!bk^{\alpha _{2}}\!-\!2g)%
}{2\sqrt{\left( ak^{\alpha _{1}}+bk^{\alpha _{2}}\right) ^{2}-4g\left(
ak^{\alpha _{1}}+bk^{\alpha _{2}}\right) }}.  \label{group-vel}
\end{equation}%
%
%
%
%
%
%
%
%
%
%
%
%
%
%
%
%
%
%
%
%
%
%
%
%
%
%
%\begin{widetext}
%\begin{equation}
%w'(k)=\frac{1}{2}\frac{\left(a\alpha_1k^{\alpha_1-1}+b\alpha_2k^{\alpha_2-1}\right)
%(a|k|^{\alpha_1}+b|k|^{\alpha_2}-2g)}
%{\sqrt{\left(a|k|^{\alpha_1}+b|k|^{\alpha_2}\right)^2-4g\left(a|k|^{\alpha_1}+b|k|^{\alpha_2}\right)}}.
%\end{equation}
%\end{widetext}
For $a=1/4,\,b=3/4,\,\alpha _{1}=1,\,\alpha _{2}=2,\,g=1$, the minimum of
expression (\ref{group-vel}) is $\left( w^{\prime }(k)\right) \!_{\min
}\approx 2.454$. To verify this prediction, we simulated Eq. (\ref{MUFNLS})
with the initial input
\begin{equation}
\psi (x,0)=1+i\cos (\sqrt{2}x)e^{-x^{2}}.  \label{Ini}
\end{equation}%
The simulations have produced the boundary of the area covered by the
propagating wave perturbations, as shown in Fig.~\ref{MI}(b1), \textit{viz}.,%
\begin{equation}
x=\pm w_{\text{min}}^{\prime }(k)t,\quad w_{\text{min}}^{\prime }(k)\approx
2.454,  \label{boundary}
\end{equation}%
which obviously agrees with the prediction.

\item {} Second, we change $a=1/4$ to $a=-1/4$, keeping values of the other
parameters. In this case, the system may feature both normal (positive) and
anomalous (negative) diffractions. The corresponding MI region in the $%
(k,\alpha _{2})$ plane is plotted in Fig.~\ref{MI}(a2). Compared to the
result shown in Fig.~\ref{MI}(a1), the MI gain is more sensitive to the
value of $\alpha _{2}$. The result of the simulated propagation with the
same input (\ref{Ini}) and $\alpha _{2}=2$ is exhibited in Fig.~\ref{MI}%
(b2), where the red lines denote the same boundary as given by Eq. (\ref%
{boundary}), in agreement with the fact that
\bee
w_{\min}^{\prime }(k)\approx 2.454
\ene
 is valid in the present case too.

\item {} Third, we consider the MI for parameters $a=-1/4,b=3/4,\alpha
_{1}=1.5,g=1$. In this case, Eq. (\ref{g>0}) yields
\begin{equation}
0<3|k|^{\alpha _{2}}-|k|^{3/2}<16,  \label{MI3}
\end{equation}%
and the maximum MI gain is attained at $3|k|^{\alpha _{2}}-|k|^{3/2}=8$. It
is observed in Fig.~\ref{MI}(a3) that the respective MI\ bandwidth first
expands and then shrinks, in contrast with the previous cases. The simulated
propagation initiated by the same input (\ref{Ini}) which was used above,
with $\alpha _{2}=1$, is exhibited in Fig.~\ref{MI}(b3), where the red lines
denote the boundary produced by
\begin{equation}
x=\pm w_{\text{min}}^{\prime }(k)t,\quad w_{\text{min}}^{\prime }(k)\approx
0.339,  \label{boundary3}
\end{equation}%
in agreement with the prediction of Eq. (\ref{group-vel}).

\item {} Finally, we set $a=-1/4,b=3/4,\alpha _{1}=2,g=-1$. In this case,
Eq. (\ref{g<0}) predicts the MI in the region of
\begin{equation}
-16<3|k|^{\alpha _{2}}-|k|^{2}<0.  \label{MI4}
\end{equation}%
Note that, while the NLS or FNLS equation with the single diffraction term
and self-defocusing nonlinearity ($g=-1$) does not give rise to MI, here MI
takes place in the two-LI FNLS model with $g=-1$ and opposite signs of the
two diffraction coefficients. The respective MI\ region in the $(k,\alpha
_{2})$ plane is plotted in Fig.~\ref{MI}(a4). For this case, the propagation
of the oscillatory perturbations initiated by input (\ref{Ini}), with $%
\alpha _{2}=1$, is displayed in Fig.~\ref{MI}(b4), where the red line is the
boundary predicted by Eq. (\ref{group-vel}) in the form of
\begin{equation}
x=\pm w_{\text{min}}^{\prime }(k)t,\quad w_{\text{min}}^{\prime }(k)\approx
1.139.  \label{boundary4}
\end{equation}%
\end{itemize}

\begin{figure}[!t]
\centering
\vspace{-0.15in} {\scalebox{0.8}[0.8]{\includegraphics{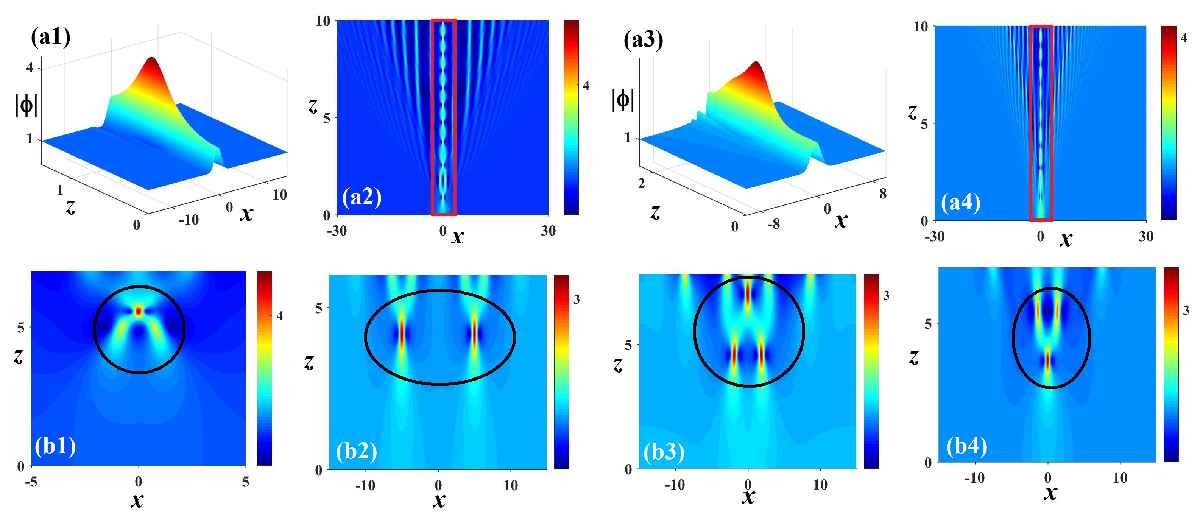}}}\hspace{-0.3in}
\vspace{0.1in}
\caption{{\protect\small (a1,a2) The excitation of the first-order RW by
input (\protect\ref{gau1}) and subsequent propagation. The system's
parameters are $(a,b,\protect\alpha _{1},\protect\alpha %
_{2},g)=(1/4,3/4,1.0,2.0,1)$. (a3,a4) The excitation of the first-order RW b
y input initial condition (\protect\ref{gau2}) and the subsequent
propagation. The system parameters are $(a,b,\protect\alpha _{1},%
\protect\alpha _{2},g)=$$(-1/4,3/4,2.0,1.0,-1)$. (b1-b4) The excitation of
different types of second-order RWs by input (\protect\ref{gau2}). The
system parameters are $(a,b,\protect\alpha _{1},\protect\alpha %
_{2},g)=(1/4,3/4,1.0,2.0,1)$. }}
\label{RW1}
\end{figure}

\begin{table}[!t]
\caption{The peak value and propagation distance at which it appears for
different LIs $\protect\alpha _{2}$. Other parameters are fixed as $(a,b,%
\protect\alpha _{1},g)=(1/4,3/4,1.0,1)$. \protect\vspace{0.05in}}\centering
\begin{tabular}{c|cccc}
\hline
Item & $\alpha_2=1.00$ & $\alpha_2=1.46$ & $\alpha_2=1.70$ & $\alpha_2=2.00$
\\ \hline
Peak value & 7.543 & 9.364 & 6.408 & 4.536 \\ \hline
Emergence propagation distance & 1.078 & 0.836 & 0.828 & 0.852 \\ \hline
\end{tabular}%
\label{table1}
\end{table}

\subsection{Formation of rogue waves}

To address the excitation of RWs on top of the CW, we consider solutions of
Eq. (\ref{MUFNLS}) with the focusing nonlinearity ($g=1$) and initial
conditions in the form of a linear superposition of the CW and $N$ Gaussian
perturbations~\cite{Ga20},
\begin{equation}
\psi (x,0)=1+\sum_{j=1}^{N}c_{j}\exp \left[ -\left( x-x_{j}\right)
^{2}/v_{j}^{2}\right] ,  \label{RWI}
\end{equation}%
where $c_{j},\,x_{j},\,v_{j}$ are the amplitude, central coordinate, and
width for the $j$-th perturbation term, respectively.

Systematic simulations make it possible to make the following conclusions
about the RWs:

\begin{itemize}
\item {} The formation of the first-order RW was addressed first. In this
case, the initial condition amounts to
\begin{equation}
\psi (x,0)=1+\exp {(-x^{2})},  \label{gau1}
\end{equation}%
i.e., expression (\ref{RWI}) with $N=1,\,x_{1}=0$, and$\,w_{1}=1$. First, we
consider the case of $a=1/4,\,b=3/4,\alpha _{1}=1,\,\alpha _{2}=2$ in Eq. (%
\ref{MUFNLS}). The profile of the RW excited in this case is displayed in
Fig.~\ref{RW1}(a1), with the peak value $\mathrm{max}(|\psi |)\simeq 4.536$,
which emerges at $z=0.85$. This value is much higher (by a factor $\approx 3$%
) than the integrable NLS equation generates from the same input. The peak
value and the propagation distance at which it appears is presented in
Table~1 for different L\'{e}vy indices $\alpha _{2}$. It is found that the
peak value increases when $\alpha _{2}\in \lbrack 1,1.46]$, and then
decreases. The largest value, $\mathrm{max}(|\psi |)\simeq 9.364$, is
attained at $\alpha _{2}=1.46$, being three times higher than its
counterpart [$\mathrm{max}(|\psi |)=3$] produced by the integrable NLS
equation. The evolution of the RW pattern at later times is exhibited in
Fig.~\ref{RW1}(a2), being similar to dynamics of a breather. Note that RWs
can also be excited in the case of the defocusing nonlinearity ($g=-1$) and
opposite signs of the two diffraction coefficients. For instance, this
happens in the case of $\alpha _{1}=2,\,\alpha _{2}=1$ and $a=-1/4,\,b=3/4$.
The respective amplitude evolution plot is shown in Fig.~\ref{RW1}(a3),
where the peak value is $\mathrm{max}(|\psi |)\approx 3.925$, which emerges
at $z=1.264$. In this case, later propagation again exhibits the appearance
of a breather, see Fig.~\ref{RW1}(a4).

\begin{figure}[!t]
\centering
\vspace{-0.15in} {\scalebox{0.8}[0.8]{\includegraphics{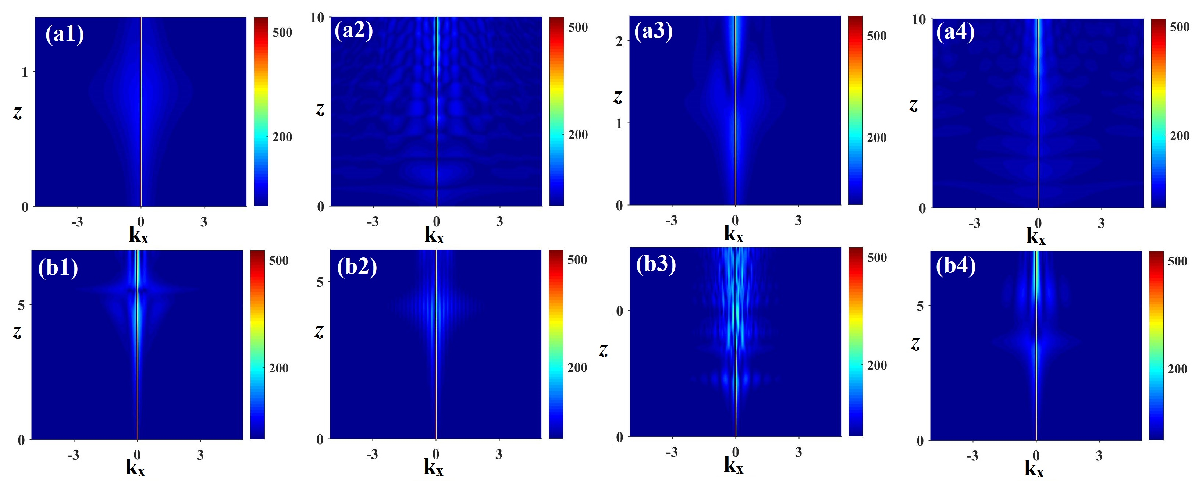}}}\hspace{-0.3in}
\vspace{0.1in}
\caption{{\protect\small The propagation of RWs from Figs.~\protect\ref{RW1}%
(a1-b4) in the Fourier space, with wavenumber }$k_{x}${\protect\small \
conjugate to coordinate }$x$. {\protect\small Parameters are same as in
Figs.~\protect\ref{RW1}(a1-b4).}}
\label{Fo}
\end{figure}

\item {} Next, we address the excitation of second-order RWs. In this case,
input (\ref{RWI}) with $N=2$ amounts to
\begin{equation}
\psi (x,0)=1+\sum_{j=1}^{2}c_{j}\exp \left[ -\left( x-x_{j}\right)
^{2}/v_{j}^{2}\right] .  \label{gau2}
\end{equation}%
We consider the formation of such RWs under the action of the focusing
nonlinearity ($g=1$) with parameters $a=1/4,\,b=3/4,\alpha _{1}=1,\,\alpha
_{2}=2$ in Eq. (\ref{MUFNLS}). Different types of the second-order RWs can
be produced by taking appropriate values of $x_{1,2}\,$and fixed $%
c_{1,2}=0.1,\,v_{1,2}=2.5\,$ in the input given by Eq.~(\ref{gau2})~\cite%
{Ga20}. As displayed in Fig.~\ref{RW1}(b1), we obtain a clustered
second-order RW, choosing $x_{1}=-x_{2}=1.6$. The second-order RW with a
split shape is generated by $x_{1}=-x_{2}=5$, see Fig.~\ref{RW1}(b2).
Further, the choice of $x_{1}=-x_{2}=2$ produces the triplet RW, see Fig.~%
\ref{RW1}(b3). A flipped triplet RW, observed in Fig.~\ref{RW1}(b4) is
produced by $x_{1}=-x_{2}=1$. The peak values for these four types of the
second-order RWs are $\mathrm{max}(|\psi |)\approx $ ${5.171,\,3,419,%
\,3.358,\,3.831}$, respectively.
\end{itemize}

In addition to Figs.~\ref{RW1}(a1-b4), the $z$-evolution of Fourier
transform $\hat{\psi}(k_{x},z)$ of $\psi (x,z)$ is displayed in Figs.~\ref%
{Fo}(a1-b4).

Thus, it is concluded that the fractional diffraction terms in Eq.~(\ref%
{MUFNLS}) has a significant impact on the MI and formation of RWs. As a
result, it is possible to select proper parameters for the excitation of RWs
of different orders in the framework of FNLS equation (\ref{MUFNLS}) with
two different fractional-diffraction terms.

\section{Conclusions and discussions}

We have proposed %a novel model, namely,
the two-LI\ (two-L\'{e}vy-index)
FNLS (fractional nonlinear Schr\"{o}dinger) model, which includes two
diffraction terms with different LIs $\alpha _{1,2}$. Experimentally, this
system can be built as a fiber laser with two fractional dispersions (or
diffractions) provided by two properly designed holograms which emulate
phase shifts corresponding to the fractional terms. If the term with $\alpha
_{2}>1$ is added to one with $\alpha _{1}=1$, which is the critical value of
the LI that, in the combination with the cubic self-focusing nonlinearity,
gives rise to the degenerate family of unstable TSs (Townes solitons), the
degeneracy is lifted, and the soliton family is stabilized. The MI
(modulational instability) is also investigated in the framework of the
present model. In particular, the MI is possible even in the case of the
defocusing nonlinearity if the two diffraction terms appear with opposite
signs. Furthermore, the first- and second-order RWs (rogue waves) are
constructed by means of directs simulations of the underlying FNLS equation.

As concerns directions for the extension of the analysis reported in this
paper, it will be relevant to consider elastic interaction of two or several stable
solitons generated from the two-LI model modulated by an external potential.
A challenging possibility is to test
stabilization of solitons which, in the case of the
single-fractional-diffraction term with $\alpha <1$, are subject to the
supercritical collapse \cite{Ch18,Qiu20}, by means of the additional
fractional-diffraction term. Further, the analysis can be extended for FNLS
equations with more than two different fractional-diffraction terms. Another
challenging issue is if the two-LI scheme can be implemented in the
two-dimensional geometry, with $\left( -\partial ^{2}/\partial x^{2}\right)
^{\alpha _{1,2}/2}$ replaced by $\left( -\nabla ^{2}\right) ^{\alpha
_{1,2}/2}$, see Eq. (\ref{FNLS-0}). And one may can consider fractional-diffraction terms with unequal LIs acting in two transverse directions, e.g., $a\left(-\partial ^{2}/\partial x^{2}\right)
^{\alpha _{1}/2}+b\left(-\partial ^{2}/\partial y^{2}\right)
^{\alpha _{2}/2}$ with $\alpha_1\not=\alpha_2$, whether anisotropic stable solitons (including fundamental and vortex ones) can be found. The stabilization of  TSs in fractional two-dimensional setting is also an important issue to be further considered in future. 

%\vspace{0.2in}\noindent \textbf{Acknowledgments} %\vspace{0.1in}

%The authors would like to thank the referees for their valuable comments and
%suggestions that have substantially improved our manuscript.
%This work was supported by the National Natural Science Foundation of China, grant No.
%11925108 and 12001246, and Israel Science Foundation, grant No. 1695/22.

\vspace{0.25in}
{\small\noindent

\noindent {\bf Ethics.} This article does not contain any studies with human or animal subjects.

\noindent {\bf Data accessibility.} This article has no additional data.

\noindent {\bf Declaration of AI use.} We have not used AI-assisted technologies in creating this work.

\noindent {\bf Authors' contributions.} M.Z.: Conceptualization, Methodology, Investigation, Analysis, Writing-original draft.
Y.C.: Conceptualization, Analysis, Writing-reviewing and editing. Z.Y.: Conceptualization, Methodology, Formal Analysis, Supervision, Funding acquisition, Writing-reviewing and editing. B.A.M.: Conceptualization, Analysis, Funding acquisition, Writing-reviewing and editing.

\noindent {\bf Competing interests.} We declare we have no competing interests.

\noindent {\bf Funding.} This work was supported by the National Natural Science Foundation of China, grant No.
11925108 and 12001246, and Israel Science Foundation, grant No. 1695/22. }

\end{document}